\documentclass[smallextended]{svjour3}       
\smartqed  
\usepackage{color}
\usepackage{graphics}
\usepackage{graphicx}
\usepackage{amssymb}
\usepackage{epstopdf}
\usepackage{mathrsfs}
\usepackage{amsmath}
\usepackage[title]{appendix}
\journalname{Journal of Low Temperature Physics}
\begin{document}

\title{Information entropy for a two-dimensional rotating Bose-Einstein condensate}

\author{R. Kishor Kumar$^1$, B. Chakrabarti$^1$  \and A. Gammal$^1$}

\institute{1 : 
           Instituto de F\'{i}sica, Universidade de S\~{a}o Paulo, 05508-090 S\~{a}o Paulo, Brazil \\
              \email{kishor@if.usp.br} }

\date{Received: date / Accepted: date}
\maketitle

\begin{abstract}
We study the information entropy, order, disorder, and complexity for 
 the two-dimensional (2D) rotating and nonrotating Bose-Einstein condensates. 
The choice of our system is a complete theoretical laboratory where the complexity is controlled by 
the two-body contact interaction strength and the rotation frequency ($\Omega$) of 
the harmonic trap. The 2D nonrotating condensate shows the complexity of the category I where 
the disorder-order transition is triggered by the interaction strength. In the rotating condensates, 
 $\Omega$ is chosen as the disorder parameter when the interaction 
strength is fixed. With respect to $\Omega$, the complexity shifts
between maximum and minimum confirm the existence of category II complexity in the rotating condensate.
Also, We consider the interaction strength as the 
disorder parameter when $\Omega$ is unchanged and complexity as 
a function of interaction strength exhibits category III complexity. 
The present work also includes 
the calculation of upper bound and lower bound of entropy for 2D quantum systems.

\end{abstract}
\section{Introduction}
\label{introduction}
Information theory plays an important role in the study of quantum systems, and it has been
 successfully used in the analysis of electron densities in atoms and molecules~\cite{Gadre1979,Sears1981,Koga1983,Guevara2003,Guevara2005,Sagar2005,Sen2005}.
The information theoretic approach of entropy maximization was applied in the 
analysis of Compton profiles and electron momentum distributions~\cite{Gadre1979,Sears1981}.
Also, the interpretation of quantum information theory is essential for quantum optics and 
condensed matter physics in the information transmission and computation~\cite{inf-opt}. 
The measurements of the observable in quantum experiment help to analyze the quantum systems and 
 provide information about the state of the system. Importantly measurements of 
entropy may help to identify the non-equilibrium state of the quantum system. 
 
The universal trend of the information entropy both for fermions and bosons is an 
important observation for the study of quantum mechanical 
systems~\cite{Massen1998,Massen2001,Massen2002}. Information entropy is
calculated using one-body density in position space ($S_r$) and in momentum 
space ($S_k$) obeys same approximate functional form 
$\sim$ $a$+$b$$N^{1/3}$, (where $N$ is the number of particles) universally for all 
types of quantum many-body systems. The net information entropy is also 
an increasing function of $N$. A simple functional form $S = a+b\ln N$ 
holds approximately for atoms, nuclei, atomic clusters, and 
correlated bosons in a trap. An important step in this direction is the 
discovery of entropic uncertainty relation (EUR) for a three-dimensional (3D) system 
$S_r+S_k~\geqslant ~3(1+ \ln \pi)~\cong 6.434~(\hbar =1)$~\cite{Bial1975,Guevara2003,Sagar2005}. 
The lower limit is attained for the noninteracting model when the distribution is Gaussian. 
A direct connection between information entropy and kinetic energy for the quantum many-body system is also 
established~\cite{Gadre1985,Gadre1987,Chat2005}. 
The total entropy is an increasing function of the 
number of particles in the system, independently of whether the system is an 
atom or a nucleus (as in references~\cite{Gadre1985,Gadre1987}). 
Recently it has been observed that for K-shell electrons of atoms 
the total correlated entropy decreases 
as one goes along the periodic table~\cite{Saha2017,Sekh2018}.  
However, order, disorder, and complexity are the three important measures 
which are inherently connected with the measure of entropy. 
We note that concepts of entropy and disorder are decoupled 
in the most of the applications~\cite{SDL-measure,Massen1998,Massen2001,Massen2002,Sriram}. 
When entropy increases, the Landsberg order parameter also increases 
and it was concluded that simultaneous increase in entropy and order can be 
explained if entropy and disorder are decoupled~\cite{N1,N2}. It has also been explicitly shown for $N$ 
trapped dipolar Bose gas that order and entropy increase simultaneously~\cite{Sriram}. 
The recent theoretical observation explained that the total entropy of the Bose gas in the 
3D trap is associated with the atoms in an excited state, although the entropy of 
particles in the ground state is non-zero~\cite{Kim2018}.

The investigation of rotating Bose-Einstein condensate is one of the central 
topics in the study of ultracold quantum physics due to its interesting features 
that includes an array of orderly aligned lattices in the quantum-Hall regime, 
Tkachenko oscillations in the lowest Landau level, bending of 
vortex lines, and so on, which can be traced by several review papers 
and book on the subject (as references in \cite{Review-vortex,BEC-Books}). 
The vortices are observed in experiments by quantum engineering techniques based on the 
atom-field coupling~\cite{Matthews1999}, topological phase 
manipulation~\cite{Leanhardt2002}, with synthetic magnetic fields~\cite{Lin2009}, 
and rotating the magnetic trap~\cite{Madison}.
Following the experiments, several theoretical investigations have been made to study the properties of 
BECs including vortices~\cite{feder,kishor-vort,bao}. 
In particular, the imaginary-time propagation method is used to generate the stationary 
vortices~\cite{feder}. Also, studies beyond mean-field are carried out to observe 
the fragmentation due to rotation using multi-configurational Hartree method 
for bosons~\cite{Marios2015}. 

Complexity measure is an ideal quantity that can serve as an ideal 
parameter to quantify the complex behavior of the 
different quantum systems. Various definitions of complexity exist in the 
literature~\cite{SDL-measure,N3,N4,Angu2008}. 
Two simple measures of complexity are L\'opez-Ruiz, Mancini, and Calbet (LMC)~\cite{LMC1995} and 
Shiner, Davison, and Landsberg (SDL)~\cite{SDL-measure}. 
However, the alternative definition of 
complexity $\Gamma_{\alpha\beta}$ is defined by SDL measure which is 
based on the appropriately defined notions of order and 
disorder~\cite{SDL-measure}. SDL defined the order 
parameter $\Lambda = 1- \frac{S}{S_{\mbox{max}}}$, where 
$S$ is the total information entropy and ${S}_{\mbox{max}}$ is 
the maximum entropy accessible to the system. 
At $\Lambda=0$ the system is at maximum accessible 
entropy with $S={S}_{\mbox{max}}$ and exhibits 
completely disordered state. On the other hand, $\Lambda=1$ implies 
that the system is at zero entropy and corresponds to perfect ordered state. 
For a realistic system, $\Lambda$ lies between zero and one.  
The SDL measure $\Gamma_{\alpha\beta}$ broadly classifies three categories of 
complexity as a function of the disorder~\cite{SDL-measure,kdsen}. In category I, complexity 
is a monotonically increasing function of the disorder. 
In category II, complexity is minimum both for perfect order and 
perfect disorder, exhibits a maximum at the intermediate level of disorder. 
In the category III, complexity is a monotonically decreasing function of 
the disorder. However, we follow the simplest measure of complexity 
$\Gamma=\Delta(1-\Delta)$~\cite{SDL-measure}. 

In this paper, we calculate the information entropy, order, disorder, and complexity for 
 the 2D rotating and nonrotating BECs. The justification of the choices 
 of the system: (a) it is an experimentally 
achievable highly complex system; (b) the interaction strength and the 
rotational frequency can both serve as a disorder parameter; and (c) higher 
rotation frequencies  may lead to the system to nonequilibrium when 
many vortices are developed. Thus, this is the most attractive test bed for 
studying the complexity, order-disorder transition and also to justify whether the 
usual thermodynamical picture will be valid, i.e., order and entropy are 
coupled. It facilitates two separate phases for the study of 
 complexity. In the first phase, we considered the nonrotating 
 condensates and observed the category I complexity~\cite{SDL-measure}. 
In the second phase, the rotating condensate is considered 
 to study the categories of complexity. The rotation frequency is considered as  
 a disorder parameter when the interaction strength is unchanged. 
The complexity for rotating condensate is minimum both for perfect order 
and disorder, also, it exhibits a hump at some 
critical rotation frequency ($\Omega=\Omega_c$) evidences a category 
II type of complexity in the rotating condensate~\cite{SDL-measure}. 
For the rotating condensate with the fixed rotation frequency,  
complexity has now decreased with 
increase in interaction strength (disorder parameter). 
It exhibits category III complexity. 
For the study of order-disorder transition, one must have all the 
fundamental relation of entropy lower bound 
and upper bound on the conjugate space in two-dimensions. So, we derive the fundamental 
inequalities for 2D quantum systems~\cite{Gadre1987}. 

The paper is organized as follows. In Sec.~\ref{secII}, we present the 2D
mean-field model for the trapped BEC under rotation, numerical methods used and 
quantities of interest of this work. In Sec.~\ref{secIII}, we report
entropy calculations for nonrotating and rotating BECs by varying the
strength of contact interaction strength and 
rotation frequency. The paper is concluded in Sec.~\ref{secIV}.
\section{Formalism}
\label{secII}
At ultra-low temperatures, the properties of a Bose-Einstein condensate of $N$ 
atoms, each of mass $m$, in rotating frame can be described by the 3D mean-field  
Gross-Pitaevskii (GP) equation~\cite{Review-vortex}. The external trapping potential 
is provided by the usual 3D harmonic trap, with a strong pancake-shaped symmetry 
and corresponding trap aspect ratio $\lambda$. 
A  strong pancake-shaped trapping potential, $V_{\it{trap}}$ is assumed to be of the form 
\begin{eqnarray}
V_{\it{trap}}({\mathbf r}) = \frac{1}{2} m \omega^2 \left( x^2+y^2+ \lambda^2 z^2 \right), \nonumber
\end{eqnarray}
where $m$ and $\omega$ are the mass and trap frequency respectively.

For the present study, we assumed pancake-shaped trap with $\lambda=10$. 
So, we reduce the corresponding 3D equation to a two-dimensional form
 by assuming the usual factorization of the wave function into the
ground state of the transverse harmonic oscillator trap and a 2D wave
function
\begin{eqnarray} \label{an2} 
\Psi({\bf r},t)\equiv  {\left(\frac{\lambda}{\pi l^2}\right)}^{1/4} 
\exp\left(\frac{-\lambda z^2}{2l^2}\right) \times \Psi_{2D}(x,y,t).
\end{eqnarray}
We performed the 2D reduction by introducing the above ansatz in the original
3D GP formalism. The final equation is in a dimensionless form where 
energy in units of $\hbar\omega$, length in units of $l=\sqrt{\hbar/(m\omega)}$,  
and time is given in units of $\tau=1/\omega$. 
The dimensionless wave-function component is given by  $\Psi(x,y,\tau) \equiv l\Psi_{2D}(x,y,t)$.
The corresponding 2D equation
\begin{eqnarray}\label{gpe2d}
 {\mathrm i} \frac{\partial \Psi_{2D}(x,y,t)}{\partial \tau} = & \biggl[-\frac{\nabla_{x,y}^2}{2}
+ V(x,y) - \Omega L_z \nonumber \\ & \, 
+g_{2D} \vert  \Psi_{2D} (x,y,t) \vert  ^2
\biggr] \Psi_{2D}(x,y,t),
\end{eqnarray}
where $V(x,y)=\frac{x^2+y^2}{2}$ is the external harmonic trap,  
$g_{2D} = 2\sqrt{2\pi\lambda}\frac{aN}{l}$ is the contact interaction parameter, 
$N$ is the number of atoms, and $a$ is the two-body atomic scattering length. 
$L_z = -{\mathrm i}\hbar(x\partial_y - y\partial_x)$ is the angular 
momentum operator with $\Omega$ the corresponding rotation frequency (in units of $\omega$).

For the numerical solution of Eq.~(\ref{gpe2d}), we employ the split-step
Crank-Nicolson method, as in Refs.~\cite{Gammal2006,CPC1}. The 
numerical simulations are carried out in imaginary time propagation on a grid with $512$ points in $x$ and $y$
directions, spatial steps $\Delta x=\Delta y=0.05$ and time step $\Delta
t=0.0005$. The wave function is renormalized to $\int dxdy|\Psi _{2D}|^{2}=1$ after each time
step. Also, the convergence of vortex solution is confirmed by conjugate gradient method~\cite{Vorst1992}. 
As found appropriate for experimentally realistic settings, 
in all the following analysis we are taking 
a pancake-shaped trap, with an aspect ratio $\lambda\,=\,10$.  

To calculate the stationary vortex states, the different initial guesses are used to check the 
convergence of ground state. From the tests, we choose the following suitable initial conditions in the
form of a combination of angular harmonics~\cite{Rokhsar}, 
\begin{equation}\label{initial}
\Psi_{2D}(x,y) = \sum_{m=0}^{L} \frac{\left(x + {\mathrm i} y \right)^m}{\sqrt{\pi (L+1) m!}} 
\exp\left[-\biggl(\frac{ x^2+y^2}{2}\biggr)  \right] \exp({\mathrm i} 2\pi \mathcal{R}_m),
\end{equation}
 where $\mathcal{R}_m$ is a randomly generated number uniformly distributed between 0 and 1, with arbitrary 
integer value for $L$ that we have considered up to $L\,=\,100$.

\subsection{Quantities of interest : Entropy, order, disorder, and complexity}
For a three-dimensional system with the continuous probability distribution  $n(\mathbf{r})$ in 
position space, the information entropy $S_r$ is calculated from
\begin{eqnarray}
S_r = -\int n(\mathbf{r}) \ln n(\mathbf{r}) d\mathbf{r},
\label{eqn:sr}
\end{eqnarray} 
where $n({\mathbf r} ) = \vert \Psi({\mathbf r})\vert^2$ is the one body density
and the corresponding information entropy in momentum space $S_{k}$ is calculated as 
\begin{eqnarray}
S_k = -\int n(\mathbf{k}) \ln n(\mathbf{k}) d\mathbf{k},
\label{eqn:sk}
\end{eqnarray} 
where $n(\mathbf{k}) = \vert{\tilde\Psi}(\mathbf{k})\vert^{2}$ is the density 
distribution in the momentum space, and the momentum space wavefunction  
$\tilde{\Psi}(\mathbf{k})$, can be obtained from the fast 
Fourier transform of $\Psi({\mathbf r})$. 
Both the density distribution 
$n(\mathbf{r})$ and $n(\mathbf{k})$ are normalized to one. 
In this case, we calculate the entropy per particle until it becomes a constant that depends 
on $\ln N$~\cite{Massen2001}. It is noted that entropy measures are scale invariant to 
the uniform change of coordinates. For the 3D system the rigorous relation 
between $S_r$ and $S_k$, total kinetic energy ($T$) and mean square radius 
has been derived using the EUR and they are 
presented by three inequalities~\cite{Gadre1987,Chat2005},  
\begin{eqnarray}
& {S_r}_{\mbox{min}} \leqslant S_r \leqslant {S_r}_{\mbox{max}}, \label{eqn:inequality1}\\
& {S_k}_{\mbox{min}} \leqslant S_k \leqslant {S_k}_{\mbox{max}},\label{eqn:inequality2}\\
& {S}_{\mbox{min}} \leqslant S \leqslant {S}_{\mbox{max}} .
\label{eqn:inequality3}
\end{eqnarray}
To calculate the above measures in two-dimensions, we need the expressions for the
lower and upper bounds like ${S_r}_{\mbox{min}}$, ${S_r}_{\mbox{max}}$, 
${S_r}_{\mbox{min}}$, ${S_k}_{\mbox{max}}$, ${S}_{\mbox{min}}$, and ${S}_{\mbox{max}}$. 
We follow the same technique from Ref.~\cite{Gadre1987} and derive the upper and lower bound entropy 
equations for 2D quantum system. These inequality relations are presented in Appendix, which are further 
utilized for the calculation of order $\Lambda = 1- \frac{S}{ {S}_{\mbox{max}}  }$, 
disorder $\Delta = \frac{S}{{S}_{\mbox{max}}}$, 
 and complexity $\Gamma = \Delta (1- \Delta)$ for the 
2D rotating and nonrotating condensates.  

\section{Results}
\label{secIII}
We start by considering a non-rotating case in Section \ref{sec:non-rotate}.
Next, we consider the rotating BEC in Section \ref{sec:rotate}.
All the following results are produced with the parameter of the contact 
interaction is given in units of the Bohr radius 
$a_0$. Adopting the length unit as $l=1.89\times 10^4 a_0$, 
the coordinates and densities are presented as dimensionless quantities. 

\subsection{Nonrotating BECs (Category I complexity)}
\label{sec:non-rotate}
We calculate the total entropy by solving the Two-dimensional 
GP Eq.~(\ref{gpe2d}) numerically for $\Omega=0$ 
and various interaction strengths $g_{2D}$. The 
information entropy $S_\rho$ for the 2D density distribution is calculated by 
\begin{eqnarray}
S_\rho = -\int n({\mathbf{\rho}}) \ln n({\mathbf{\rho}}) d{\mathbf{\rho}},
\label{eqn:srho}
\end{eqnarray} 
where $n({\mathbf{\rho}}) = \vert{\tilde\Psi_{2D}}({\mathbf{\rho}})\vert^{2}$ 
is the 2D density and ${\mathbf{\rho}}\equiv (x,y)$. 
Also, the corresponding density in momentum space $n(\mathbf{k_{\rho}})$ 
is obtained from the fast Fourier transform (FFT) and $S_{k_{\rho}}$ is calculated by using 
Eq.~(\ref{max-ent}) which is presented in Appendix. 
\begin{figure}[tbp]
\begin{center}
\includegraphics[width=0.5\textwidth]{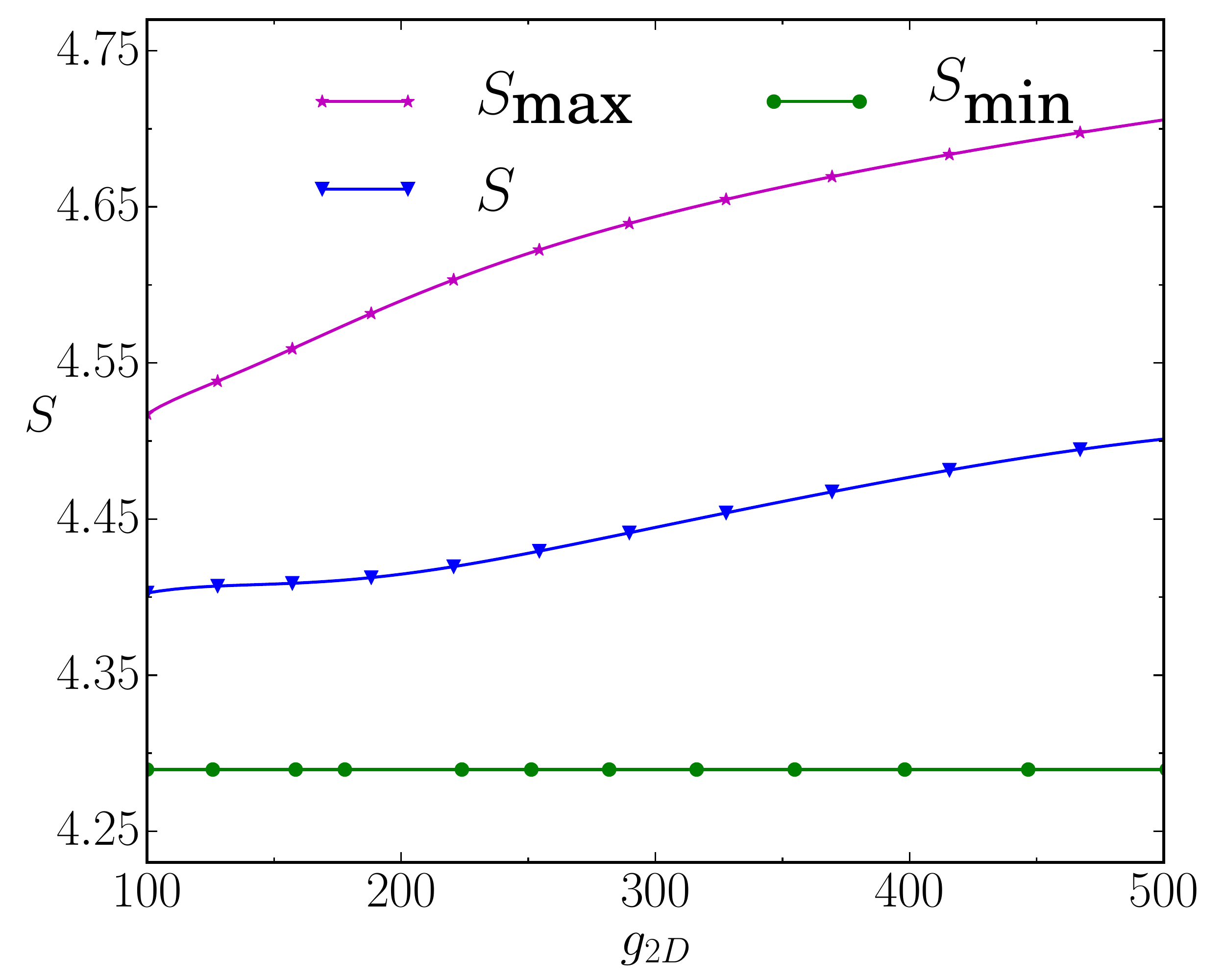}
\caption{Entropy measure of 2D nonrotating condensate. 
Upper (${S}_{\mbox{max}}$) and lower (${S}_{\mbox{min}}$) limits 
of entropy are compared with the total entropy ($S$) of the condensate 
with respect to interaction 
strengths. ${S}_{\mbox{max}}$, and ${S}_{\mbox{min}}$ are represented 
by solid lines with solid stars, triangles, and circles respectively.
 All quantities are dimensionless.} 
\label{fig1}
\end{center}
\end{figure}
The total entropy ($S=S_{\rho}+S_{k_{\rho}}$), 
upper bound (${S}_{\mbox{max}}$), 
and lower bound (${S}_{\mbox{min}}$) are plotted in Fig.~\ref{fig1}. The total entropy $S$ of the 
nonrotating condensate perfectly lies between $S_{\mbox{max}}$ and $S_{\mbox{min}}$ 
throughout the entire range of interaction strength. 
The used inequality expressions to calculate the upper and lower bounds of entropy  
are given in Appendix Eqs.~(\ref{eqn:lowupbound}). 
\begin{figure}[htbp]
\begin{center}
\includegraphics[width=0.95\textwidth]{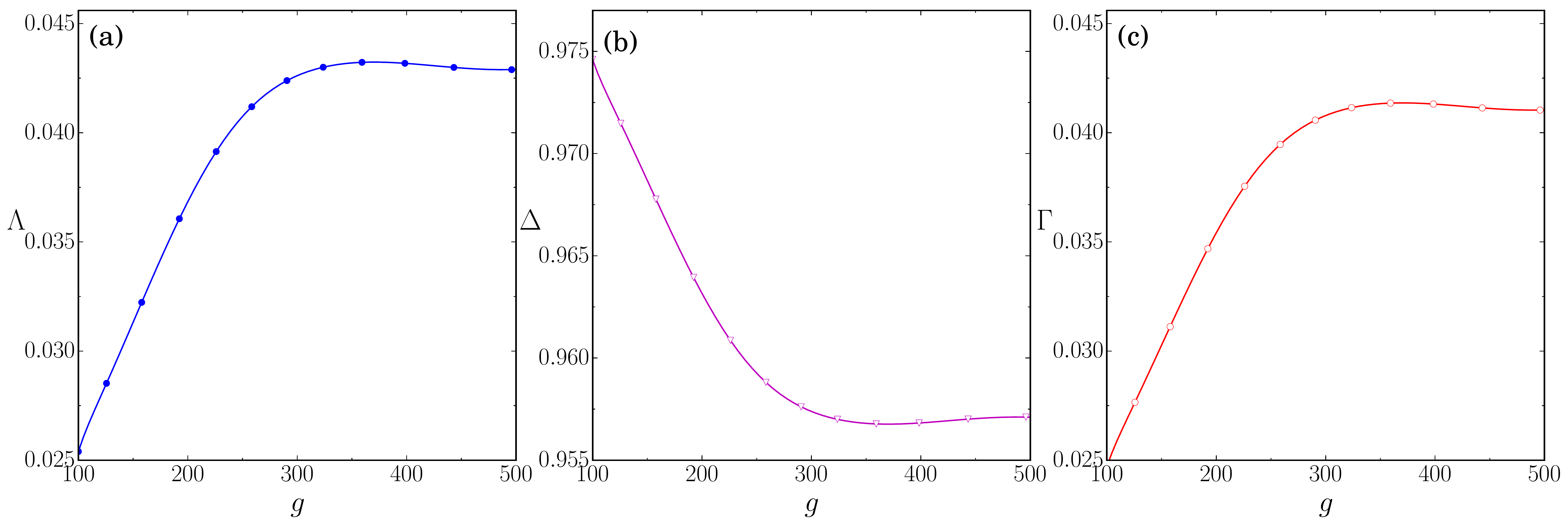}
\caption{In the upper, middle and lower panels, (a) Order, (b) disorder 
and (c) complexity of the nonrotating condensate as a function of 
interaction strengths are shown respectively. 
The smooth increase in $\Lambda$ (a) corresponding 
smooth decrease in $\Delta$ (b) and characterized category I 
complexity in (c). All quantities are dimensionless.} 
\label{fig2}
\end{center}
\end{figure} 
Further, we have calculated the corresponding Landsberg order parameter, disorder, 
and complexity in Fig.~\ref{fig2}.  
 In Fig.~\ref{fig2}(a), order increases regarding the increase in 
interaction strength $g_{2D}$ (which is proportional to a number of particles), 
and then order saturates. We observe that order and entropy both increase 
as similar to the previous observation of 3D condensate~\cite{Massen2002,Sriram}. 
This is confirming the explanation, order and entropy are decoupled. 
The corresponding measure of disorder $\Delta=(1-\Lambda)$ smoothly decreases and attains a saturation. 
From the above observations, we confirm that adding more particles to the system manifest the transition from 
disorder to order. In Fig.~\ref{fig2}(c), we plot the complexity $\Gamma=\Delta(1-\Delta)$, which increases
monotonically against disorder parameter $g_{2D}$ and then saturates. 
Thus, our nonrotating 2D condensate exhibits complexity which belongs to the 
category I~\cite{SDL-measure}. The saturation of the order, disorder, and 
complexity at a critical interaction strength $g_{2D}\approx 330$ can observe from Fig.~\ref{fig2}.

\subsection{Rotating BECs (Category II and III complexity)}
\label{sec:rotate}
In this subsection, we consider the 2D rotating BECs and investigate its entropy properties.
\begin{figure*}[tbp]
\begin{center}
\includegraphics[width=0.85\textwidth]{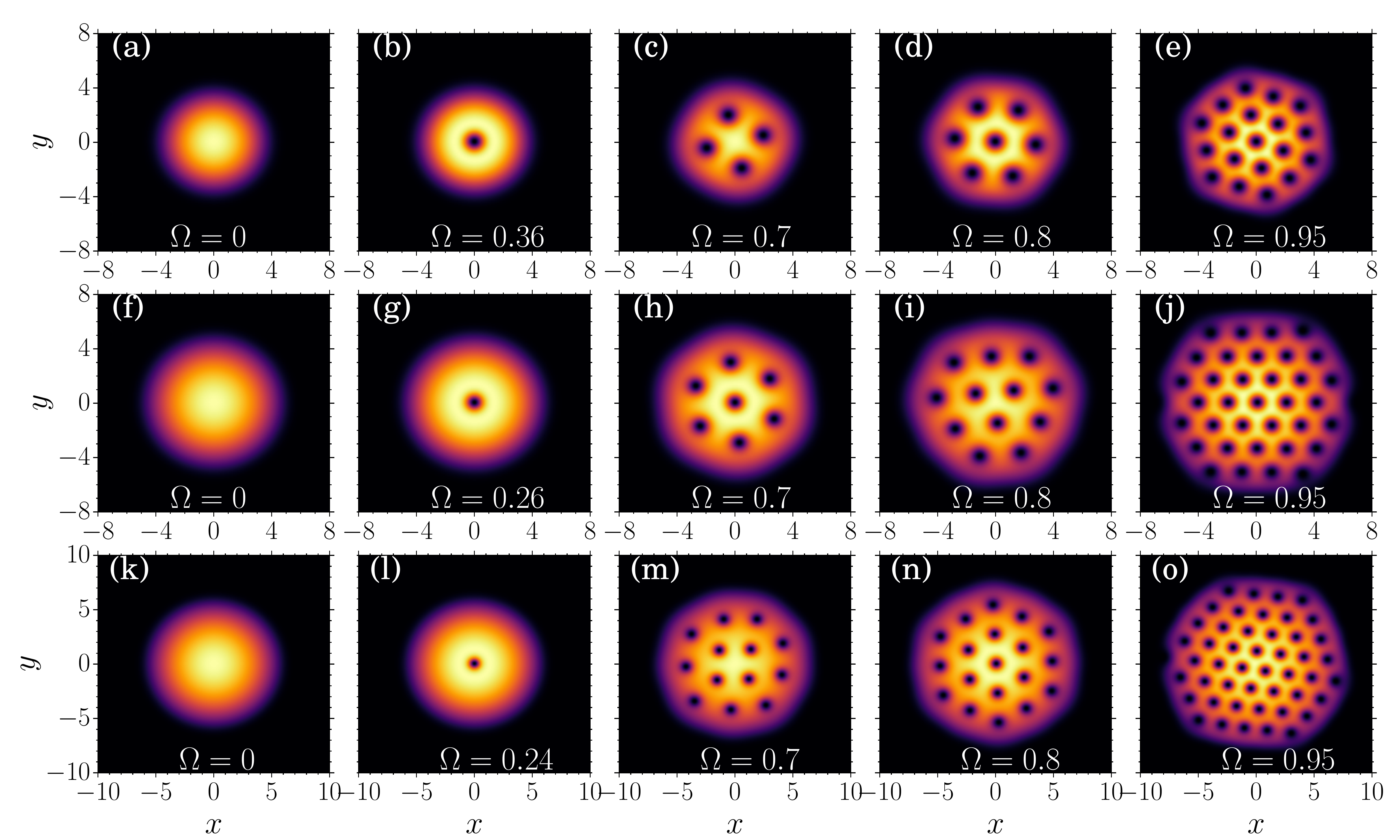}
\caption{ Two-dimensional density patterns, $|\Psi_{2D}|^{2}$  for the interaction parameters 
(a)-(e) $g_{2D}$ = 100,
(f)-(j) $g_{2D}$ = 250, and (k)-(o) $g_{2D}$ = 500. The corresponding 
rotation frequency is mentioned in each density plot.} 
\label{fig3}
\end{center}
\end{figure*}
The vortex lattice is obtained by propagating the 2D GP equation 
(\ref{gpe2d}) in imaginary time with 
non-zero rotation frequency $\Omega$. The first vortex appears at a rotation 
frequency significantly larger than critical 
frequency $\Omega_c$ for the vortex generation. The vortex lattice is strongly
influence by the trap symmetry~\cite{feder}.
\begin{figure*}[tbp]
\begin{center}
\includegraphics[width=0.95\textwidth]{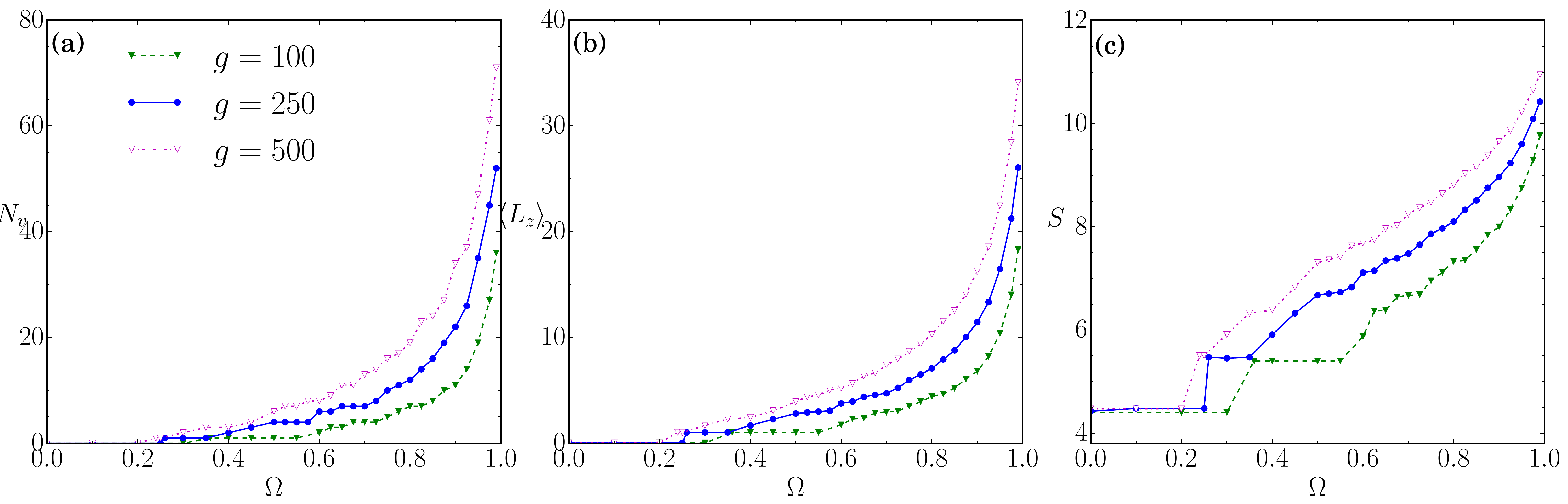}
\caption{(a) Number of vortices ($N_v$), (b) expectation value of 
angular momentum ($\langle L_z\rangle$), and (c) total entropy ($S$)
with respect to rotation frequency $\Omega$ for different interaction strengths 
$g_{2D}\,=$ 100, 250, and 500.} 
\label{fig4}
\end{center}
\end{figure*} 
In Fig.~\ref{fig3}, we display the stable solutions for densities showing the 
triangular vortex lattice for different interaction strengths and rotation frequencies.
\begin{figure}[tbp]
\begin{center}
\includegraphics[width=0.95\textwidth]{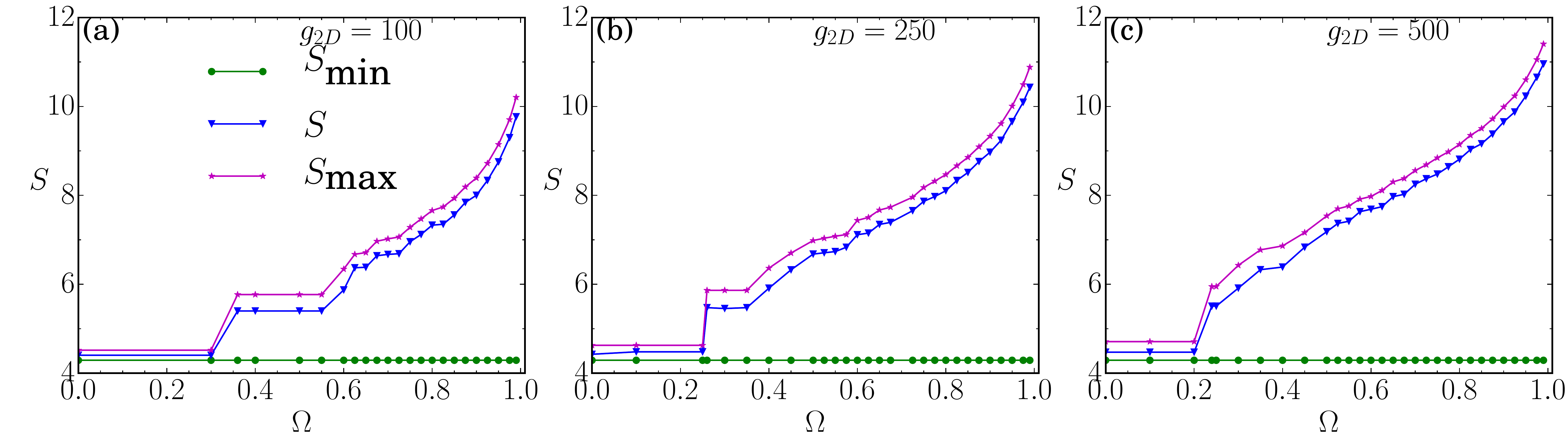}
\caption{Upper and lower limits of entropy compared with the total entropy 
of the system with respect to rotation frequency for several interaction strengths
(a) $g_{2D}$ = 100, (b) $g_{2D}$ = 250, and (c) $g_{2D}$ = 500. 
All quantities are dimensionless.} 
\label{fig5}
\end{center}
\end{figure}
Saturation in the order, disorder, and complexity is 
 the effect of the external finite sized trap. 
This critical interaction strength crucially depends on the trap aspect ratio. 

In Fig.~\ref{fig4}, we present the number of vortices ($N_v$), 
expectation value of the angular momentum of the condensate ($\langle L_z\rangle$), and total entropy ($S$) as 
a function of rotation frequency. The $N_v$, $\langle L_z\rangle$, and $S$ all three parameters are increasing 
with respect to $\Omega$. The number of vortices and angular momentum diverge at the rotation 
frequency near to harmonic trap frequency~\cite{Rokhsar}. But the entropy of 
the system increases smoothly as shown in Fig.~\ref{fig4}(c). 
There is a discontinuous transition between $N_v$ and  $\langle L_z\rangle$ both jump from zero to unity 
when the first vortex enters. Similarly, there is a jump in the total entropy, when the first vortex enters 
into the condensate. The discontinuous increase in $N_v$, $\langle L_z\rangle$ 
and $S$ with increasing $\Omega$ is due to the dynamical entry of vortices into the condensate. 
 The angular momentum essentially depends on the number of vortices, but 
it does not fluctuate significantly by the orientation of the condensate 
regarding the vortex lattice arrangement and radius of the condensate. 
So, the angular momentum does not increase significantly. 
But total entropy remains the same if the number of vortices is unchanged. 
So, between some rotation frequencies, the angular momentum goes up 
continuously while entropy may rise discontinuously. 
This difference can be visualized by comparing Figs.~\ref{fig4}\,(b) and (c).
Next, the upper limit (${S}_{\mbox{max}}$), lower limit (${S}_{\mbox{min}}$) 
of the entropy as a function of $\Omega$ is shown in Fig.~\ref{fig5}. 
Initially, total entropy $S$ lies between 
$S_{\mbox{max}}$ and $S_{\mbox{min}}$, but it diverges sharply after the vortex enters 
into the condensate, and becomes very close to the upper limit of ${S}_{\mbox{max}}$.

\begin{figure}[tbp]
\begin{center}
\includegraphics[width=0.9\textwidth]{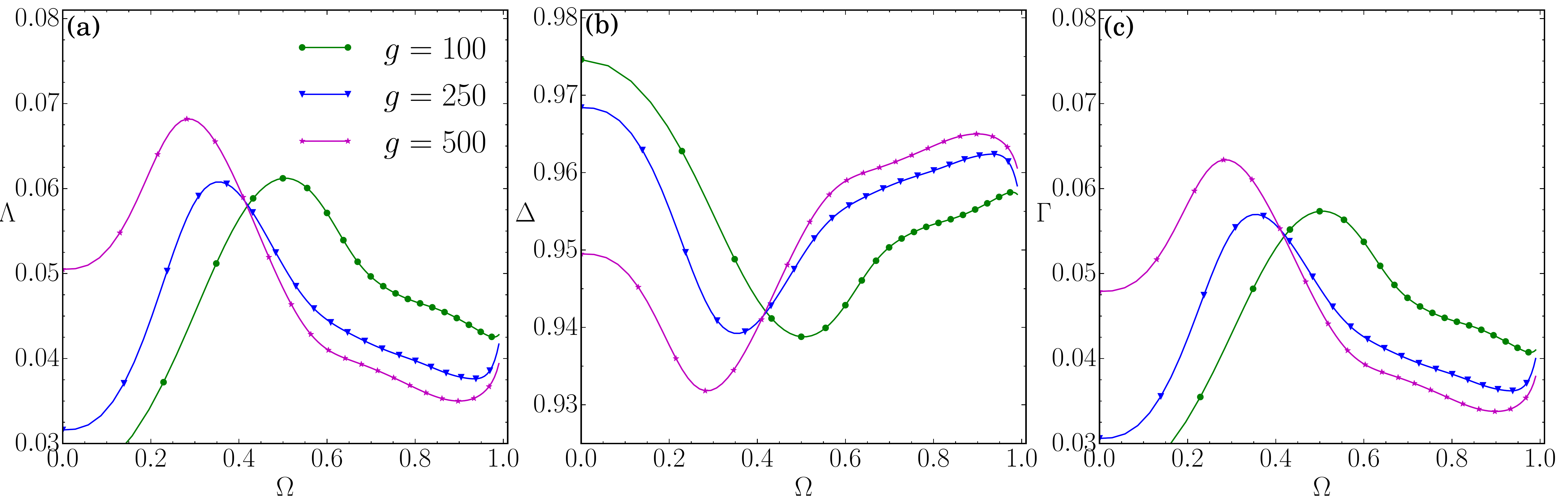}
\caption{(a) Order, (b) Disorder and (c) complexity with 
respect to rotation frequency for the BECs with different interaction strengths. 
All quantities are dimensionless.} 
\label{fig6}
\end{center}
\end{figure}

Further, we calculate $\Lambda$, $\Delta$, and $\Gamma$ for the 
rotating condensate. In the rotating condensates, rotation 
frequency plays a crucial role in determining its properties. 
The critical rotation frequency decreases monotonically 
with increasing interaction strength for rotating BECs in all trap geometries~\cite{feder}. 
 The critical rotation frequency is gradually decreased with the increasing 
$g_{2D}$, and is calculated for the interaction strengths, $g_{2D}\,=\,100, \, 250$, and 500 
are, $\Omega_c = 0.36,\, 0.26$, and 0.24 respectively. 
When the rotation frequency is significantly higher than $\Omega_c$, then 
more vortices enter into the condensate and form a triangular lattice~\cite{Review-vortex}. 
We analyze the different regimes regarding the rotation frequency
where we have no vortex or single vortex or many vortices.  
The order parameter $\Lambda$ is plotted in Fig.~\ref{fig6}(a) and shows that  
the order increases smoothly and shows a maximum value at critical rotation frequency $\Omega\,=\,\Omega_c$. 
Further, order decreases for $\Omega\,>\,\Omega_c$.
On the other hand, in Fig.~\ref{fig6}(b), disorder shows the reverse 
behavior where disorder smoothly decreases till $\Omega_c$ and then it
 increases for $\Omega\,>\,\Omega_c$. 
Here, this reverse behavior is due to 
the entry of a large number of vortices into the condensate. They adjust themselves and fills up the 
disorder. From a thermodynamics point of view, it is an unexpected result that entropy rises and disorder 
falls. However, from our previous observation of nonrotating BEC, we 
conclude that till the critical frequency, 
entropy and disorder are decoupled. Increasing the rotation 
frequency of the trap above $\Omega_c$  
increases the dissipation in the condensate. Even though more vortices 
enter at $\Omega\,>\,\Omega_c$, they are not sufficient to 
fill up the disorder. So, disorder increases smoothly as a function of $\Omega$. 
It assumes that for $\Omega\,>\,\Omega_c$, entropy and disorder are now coupled and 
it satisfies the most usual view of thermodynamics.
\begin{figure}[tbp]
\begin{center}
\includegraphics[width=0.5\columnwidth]{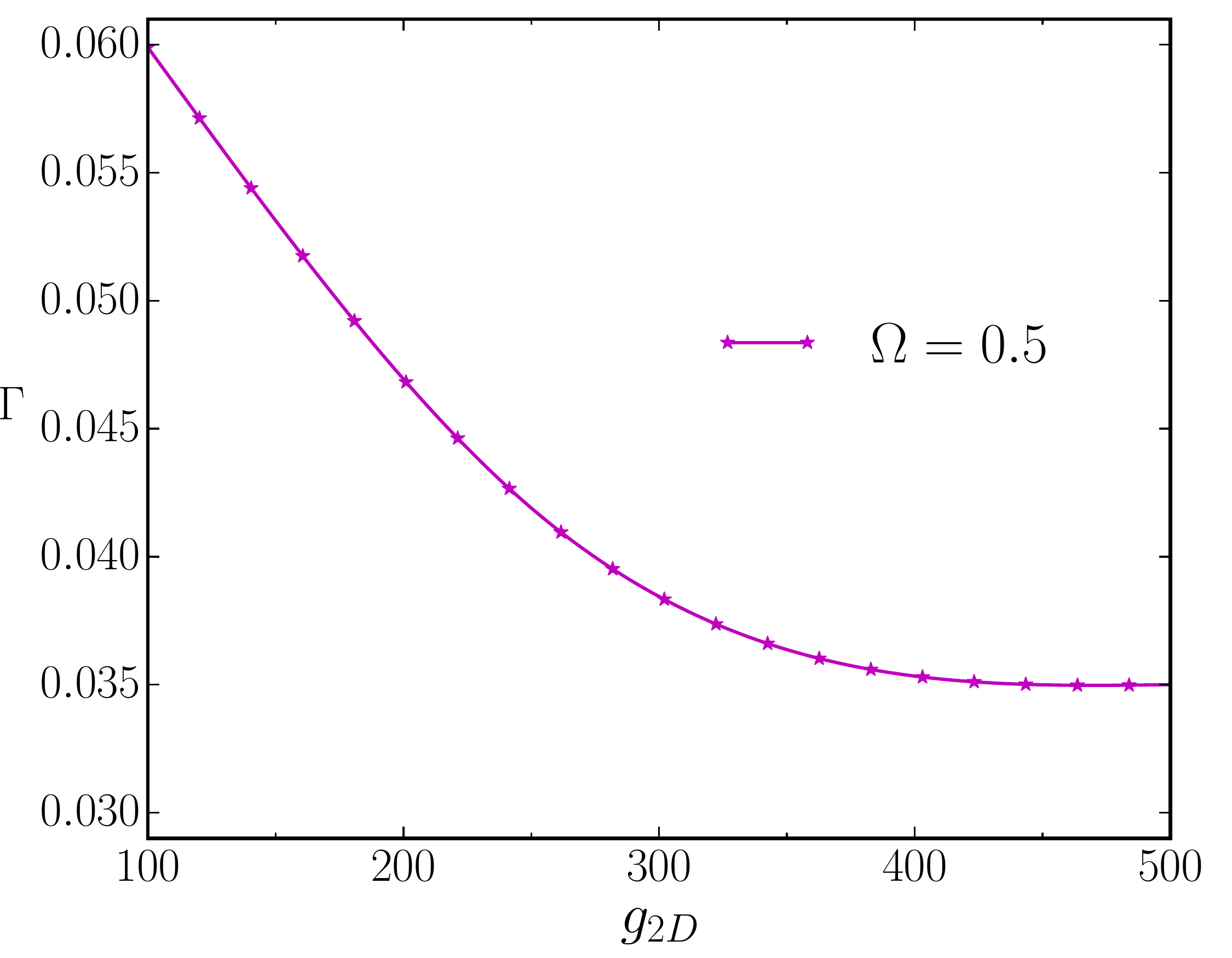}
\caption{Complexity as a function of interaction strengths show the complexity category 
III for the rotation frequency $\Omega=0.5$, which is greater than critical rotation frequency for 
the interaction strength regime presented in this plot.} 
\label{fig7}
\end{center}
\end{figure} 
In addition, we plot the complexity as a function of disorder parameter $\Omega$ in Fig.~\ref{fig6}(c). 
Complexity is minimum both for highest ordered and highest disordered 
state, but never reach to zero value. Thus, it is confirmed that the rotating condensate has
been always complex. Complexity is increased by larger rotation frequency when the condensate has a large number of vortices. Similar to order, complexity $\Gamma$ also exhibits a maximum at $\Omega_c$. 
So it confirms that the complexity belongs to category II. 
Rotating condensates exhibit two transitions. Disorder to order transition 
is continuing as far as $\Omega\,=\,\Omega_c$. In contrast, for $\Omega\,>\,\Omega_c$ 
 the system shows a transition from order to 
disorder and complexity goes down with increasing interaction strength. 
Eventually, we chose a fixed rotation frequency $\Omega$ = 0.5, which is the frequency greater than $\Omega_c$ of all the 
interaction strengths considered. The complexity as a function of disorder 
parameter $g_{2D}$  for a fixed $\Omega\,=\,0.5\,>\,\Omega_c$ is plotted in Fig.~\ref{fig7}. 
The complexity decreases with increasing interaction strength. 
We observe that the complexity goes down and the system exhibits category III complexity. 
From the above observations, we confirm the existence of 
category II and III complexities in rotating condensates. 
\section{Conclusion}
\label{secIV}
 
In this work, we have presented the calculations of information entropy, order, disorder, 
and complexity for 2D rotating and nonrotating Bose-Einstein 
condensates. In order to study the order-disorder transition, we have derived the 
fundamental inequalities of entropy lower bound 
and upper bound for 2D quantum systems. We compare the total entropy of rotating and nonrotating 
2D BECs with maximum and minimum limits of entropy. 
We observe that our system is unique which can exhibit all the three 
categories of complexity regarding SDL measure.
We have considered the two-body interaction strength as a disorder parameter 
for the observations of complexity category in nonrotating condensates. 
In general, complexity is a monotonically increasing function regarding disorder parameter in 
a category I complexity. Similarly, in nonrotating BECs, complexity increases with disorder parameter  
which confirms the existence of the category I complexity.

 Next, we have studied the entropy properties of rotating condensates. 
In the rotating condensates, we consider the rotation frequency 
as the disorder parameter when the interaction strength is fixed. 
The complexity increases until the rotation frequency reach 
$\Omega_c$ ($\Omega\,=\,\Omega_c$). Then, 
complexity starts to decrease for $\Omega\,>\,\Omega_c$. 
The condensate becomes more dissipative due to the fast rotation and the entry of many 
vortices that decrease the complexity.
This transition in maximum and minimum 
complexity shows a hump at $\Omega\,=\,\Omega_c$. 
Thus, the characteristic corresponding to category II 
complexity is satisfied by the rotating condensate.  
Finally, the rotation frequency is unchanged and the interaction strength is used as a 
disorder parameter. In this setting, the complexity goes down regarding the 
increase in disorder parameter. This kind of decreasing complexity with respect to disorder parameter 
 characterizes the existence of category III complexity in rotating condensates.

 We believe that this is the first theoretical study performed in an experimentally realizable system which exhibits all three categories of complexity. Calculation of the complexity measure 
by LMC, their comparison with SDL and finding the value of $\alpha$ and $\beta$
~\cite{SDL-measure} for three types of complexity will be the subject of the future studies.

\begin{appendices}
\section{Connection between $S_r$, $S_k$ with the total kinetic energy $T$ 
and mean square radius in two-dimensions }
\label{app:1}
Maximum value of entropy in momentum space for a 2D system is given by
\begin{eqnarray}\label{max-ent}
S_{k_{\rho}} \le -\int n(\mathbf{k_{\rho}}) \ln n(\mathbf{k_{\rho}}) d\mathbf{k_{\rho}}.
\end{eqnarray}

Dimensionless form of kinetic energy 
$T=\frac{1}{2}\int n(\mathbf{k_{\rho}}) {\mathbf{k}}_{\rho}^2 d\mathbf{k_{\rho}}$, 
where ${\mathbf{k}}_{\rho}^2\,=\, \mathbf{k}_x^2+\mathbf{k}_y^2$. We consider the density in 
momentum space $n(\mathbf{k_{\rho}})\,=\, A \exp[-\alpha {\mathbf{k}}_{\rho}^2]$, where $A$ is the 
normalization constant and $\alpha$ is the appropriate 
Lagrange multiplier. The normalization of the density with respect to 
$N$ particles is defined $\int_\infty^ {-\infty} n(\mathbf{k_{\rho}}) d\mathbf{k_{\rho}}\, =\, N$. 
It calculates  $A\,=\,\alpha N/\pi$ and $\alpha\,=\,N/2T$. 
Thus maximum value of the momentum space is given by Eq.~(\ref{max-ent}) and 
further simplification yields the maximum value of momentum space entropy is given by 
\begin{eqnarray} 
S_{k_{\rho}} \le N (1+\ln \pi) - N \ln N - N \ln \left(\frac{N}{2T}\right).
\label{Sk-max}
\end{eqnarray}
For the 2D model, we get the following relation from refs.~\cite{Bial1975,Gadre1985},
\begin{eqnarray}
S_{\rho} + S_{k_{\rho}} \ge 2 N (1+\ln \pi) - 2 N \ln N.
\label{srsk}
\end{eqnarray}
From the relations ~(\ref{Sk-max}) and (\ref{srsk}), we obtain the lower bound to $S_{\rho}$ 
\begin{eqnarray} 
S_{\rho} \ge N (1+\ln \pi) - N \ln N + N \ln \left(\frac{N}{2T}\right).
\label{Sr-min}
\end{eqnarray}
Addition of (\ref{Sk-max}) and (\ref{Sr-min}) provide the 
lower bound to the excess information entropy in the
position space over that in the momentum space.
\begin{eqnarray}
S_{\rho} - S_{k_{\rho}} \ge 2 N (1+\ln \pi) - 2 N \ln(2T).
\end{eqnarray}
Next, we calculate the upper as well as lower bounds for $S_{\rho}$ 
and $S_{k_{\rho}}$ respectively in terms of 
$\langle {{\mathbf{\rho}}}^2\rangle$  

\begin{eqnarray} 
S_{\rho} \le N (1+\ln \pi) - 2 N \ln N + N \ln \left( \langle {\mathbf{\rho}}^2\rangle \right),
\label{Sr-max}
\end{eqnarray}
where ${\mathbf{\rho}}^2=x^2+y^2$ and 
\begin{eqnarray} 
S_{k_{\rho}} \ge N (1+\ln \pi) +2 N \ln N - N \ln \left( \langle {\mathbf{\rho}}^2\rangle \right).
\label{Sk-min}
\end{eqnarray}
For density distribution normalized to unity, 
the lower and upper limits of entropy in two-dimensions took the form
\begin{subequations}\label{eqn:lowupbound}
\begin{eqnarray}
{S_{\rho}}_{\mbox{min}} &= (1+ \ln \pi) -  \ln \left( 2 T \right),  \label{subeq1} \\ 
{S_{\rho}}_{\mbox{max}} &= (1+ \ln \pi) + \ln \left( \langle {\mathbf{\rho}}^{2} \rangle \right), \label{subeq2} \\ 
{S_{k_{\rho}}}_{\mbox{min}} &= (1+ \ln \pi) - \ln \left( \langle {\mathbf{\rho}}^{2} \rangle \right), \label{subeq3} \\ 
{S_{k_{\rho}}}_{\mbox{max}} &= (1+ \ln \pi) + \ln \left( 2 T \right), \label{subeq4} \\ 
{S}_{\mbox{min}} &= 2(1+ \ln \pi), \label{subeq5} \\ 
{S}_{\mbox{max}} &= 2(1+ \ln \pi) + \ln \left(2 \langle {\mathbf{\rho}}^{2} \rangle T \right). \label{subeq6}
\end{eqnarray}
\end{subequations}
\end{appendices}
\begin{acknowledgements}
RKK, BC and AG acknowledge the support by FAPESP of Brazil under 
grants 2014/01668-8, 2016/19622-0 and 2016/17612-7,  
respectively.  AG also acknowledges the support by CNPq of Brazil.
\end{acknowledgements}

\end{document}